\documentclass[a4paper]{jpconf}
\usepackage{graphicx}
\usepackage{epstopdf}
\usepackage{amssymb}
\usepackage{lineno}
\usepackage{xspace}
\usepackage{amsmath}
\usepackage{upgreek}

\newcommand{\krl}{\ensuremath{\kern-0.18em}}
\newcommand{\krr}{\ensuremath{\kern-0.09em}}
\newcommand{\tms}{\ensuremath{\kern-0.1em\times\kern-0.2em}}

\newcommand{\ptt}{\ensuremath{p_{\mathrm{T}}}\xspace}
\newcommand{\pb}{Pb--Pb\xspace}
\newcommand{\ppb}{p--Pb\xspace}

\newcommand{\rs}[1][7~TeV]{\ensuremath{\sqrt{s}=}~#1\xspace}

\newcommand{\rsnn}[1][2.76~TeV]{\ensuremath{\sqrt{s_{\mathrm{NN}}}=}~#1\xspace}

\newcommand{\gvc}{\ensuremath{\mathrm{GeV}\krl/\krr c}\xspace}

\newcommand{\pion}{\ensuremath{\uppi}\xspace}
\newcommand{\pix}{\ensuremath{\pion^{\pm}}\xspace}
\newcommand{\pim}{\ensuremath{\pion^{-}}\xspace}
\newcommand{\pip}{\ensuremath{\pion^{+}}\xspace}

\newcommand{\km}{\ensuremath{\mathrm{K}^{-}}\xspace}
\newcommand{\kp}{\ensuremath{\mathrm{K}^{+}}\xspace}

\newcommand{\rh}{\ensuremath{\uprho^{0}}\xspace}
\newcommand{\rhm}{\ensuremath{\uprho(770)^{0}}\xspace}
\newcommand{\rhpi}{\ensuremath{\rh\krl/\pion}\xspace}
\newcommand{\rhpix}{\ensuremath{\rh\krl/\langle\pix\rangle}\xspace}

\newcommand{\ks}{\ensuremath{\mathrm{K^{*0}}}\xspace}

\newcommand{\ksm}{\ensuremath{\mathrm{K^{*}\krr(892)^{0}}}\xspace}

\newcommand{\ph}{\ensuremath{\upphi}\xspace}
\newcommand{\phm}{\ensuremath{\ph(1020)}\xspace}

\newcommand{\ksk}{\ensuremath{\ks\krl/\mathrm{K}}\xspace}

\newcommand{\phik}{\ensuremath{\ph\krl/\mathrm{K}}\xspace}

\newcommand{\pphi}{\ensuremath{\mathrm{p}\kern-0.1em/\krl\ph}\xspace}
\newcommand{\pphim}{\ensuremath{\mathrm{p}\kern-0.1em/\krl\phm}\xspace}

\newcommand{\sigs}{\ensuremath{\Sigma(1385)^{\pm}}\xspace}
\newcommand{\sigsl}{\ensuremath{\Sigma(1385)^{\pm}\kern-0.1em/\krl\Lambda}\xspace}
\newcommand{\xs}{\ensuremath{\Xi(1530)^{0}}\xspace}
\newcommand{\xsx}{\ensuremath{\Xi(1530)^{0}\kern-0.1em/\Xi^{-}}\xspace}

\newcommand{\dd}{\ensuremath{\mathrm{d}}}
\newcommand{\mpt}{\ensuremath{\langle\ptt\rangle}\xspace}

\newcommand{\dnc}{\ensuremath{\langle\dd N_{\mathrm{ch}}\kern-0.06em /\kern-0.13em\dd\eta\rangle}\xspace}
\newcommand{\dncr}{\ensuremath{\dnc^{1/3}}\xspace}

\newcommand{\raa}{\ensuremath{R_{\mathrm{AA}}}\xspace}
\newcommand{\rppb}{\ensuremath{R_{\mathrm{pPb}}}\xspace}

\newcommand{\akea}{\textit{et al}\xspace}

\begin{document}
\title{Recent hadronic resonance measurements at ALICE}

\author{A G Knospe (for the ALICE Collaboration)}

\address{Department of Physics, The University of Houston, Houston, TX, USA}

\ead{anders.knospe@cern.ch}

\begin{abstract}
In heavy-ion physics, measurements of short-lived hadronic resonances allow the properties of the hadronic phase of the collision to be studied.  In addition, resonances can be used along with stable hadrons to study parton energy loss in the quark-gluon plasma and the mechanisms that shape hadron \ptt spectra at intermediate transverse momenta.  Resonance measurements in small systems serve as a reference for heavy-ion collisions and contribute to searches for collective effects.  An overview of recent results on hadronic resonance production measured in ALICE is presented.  These results include the \ptt spectra and yields of the \rhm, \ksm, and \phm mesons in pp, \ppb, and \pb collisions at different energies as well as the \sigs and \xs baryons in pp and \ppb collisions.
\end{abstract}

Hadronic resonances, along with stable hadrons, allow the study of properties of heavy-ion collisions, both in the early (quark-gluon plasma) and late (hadronic) stages of their evolution.  Energy loss of partons in nuclear matter can be studied by measuring the nuclear modification factors \raa and \rppb.  The yields of resonances may be modified even after chemical freeze-out by re-scattering and regeneration, which are (pseudo-)elastic scattering processes expected to have their greatest strength at low \ptt ($\lesssim 2$~\gvc)~\cite{Bleicher_Stoecker,Markert_thermal,Vogel_Bleicher}.  Measurements of resonance and stable-hadron yields can be used along with theoretical models~\cite{Markert_thermal,Torrieri_thermal,Torrieri_thermal_2001b} to estimate the properties (temperature and lifetime) of the hadronic phase of heavy-ion collisions.  The various mechanisms that may determine the shapes of particle \ptt spectra, including vacuum fragmentation, quark recombination, hydrodynamic flow, re-scattering, and regeneration, can be studied through comparison of different measurements of multiple particle species (including hadronic resonances) with differing masses and quark content.  These proceedings present measurements from the ALICE experiment that are related to these topics.

\begin{figure}
\includegraphics[width=9cm]{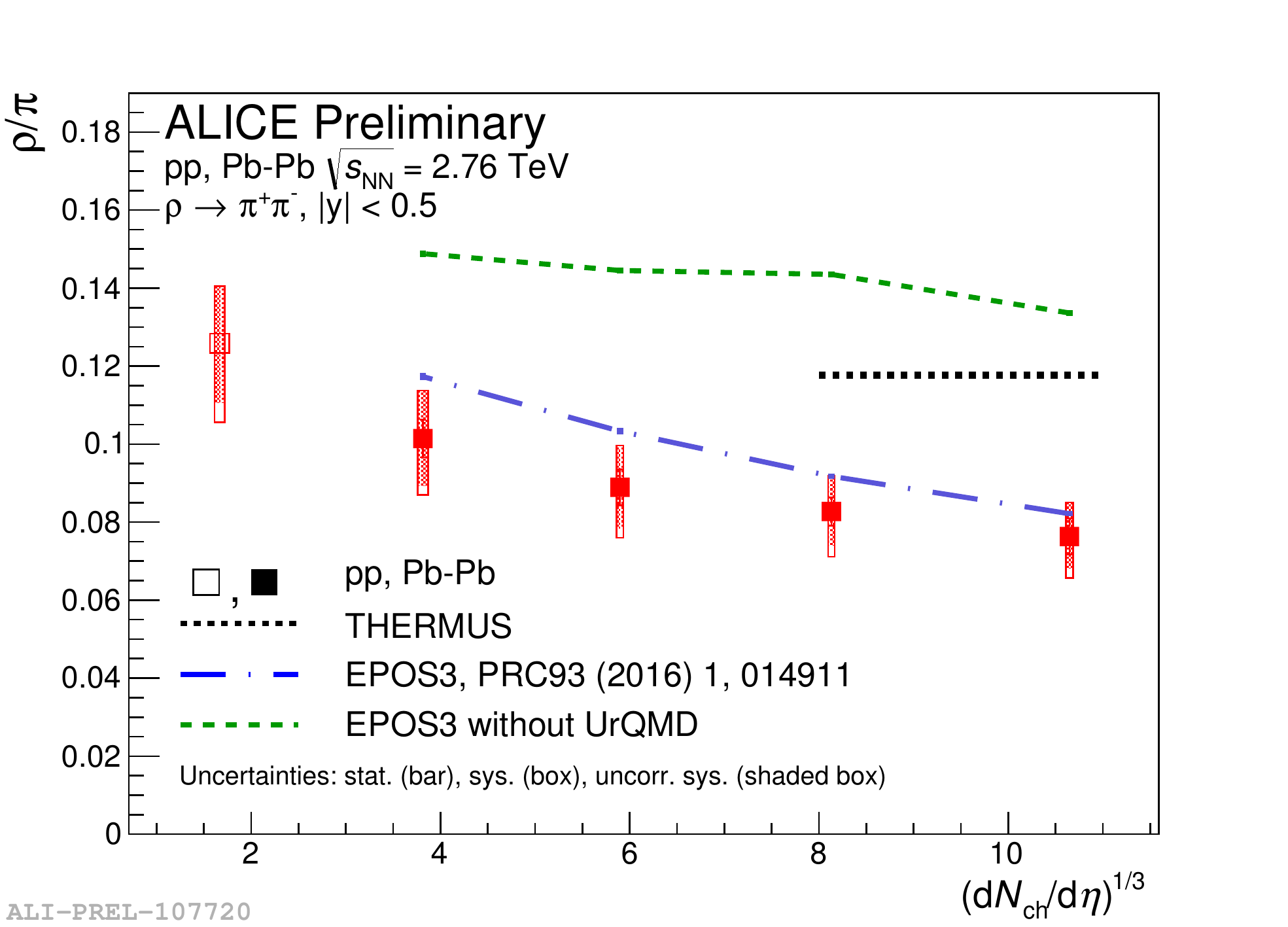}
\includegraphics[width=6.15cm]{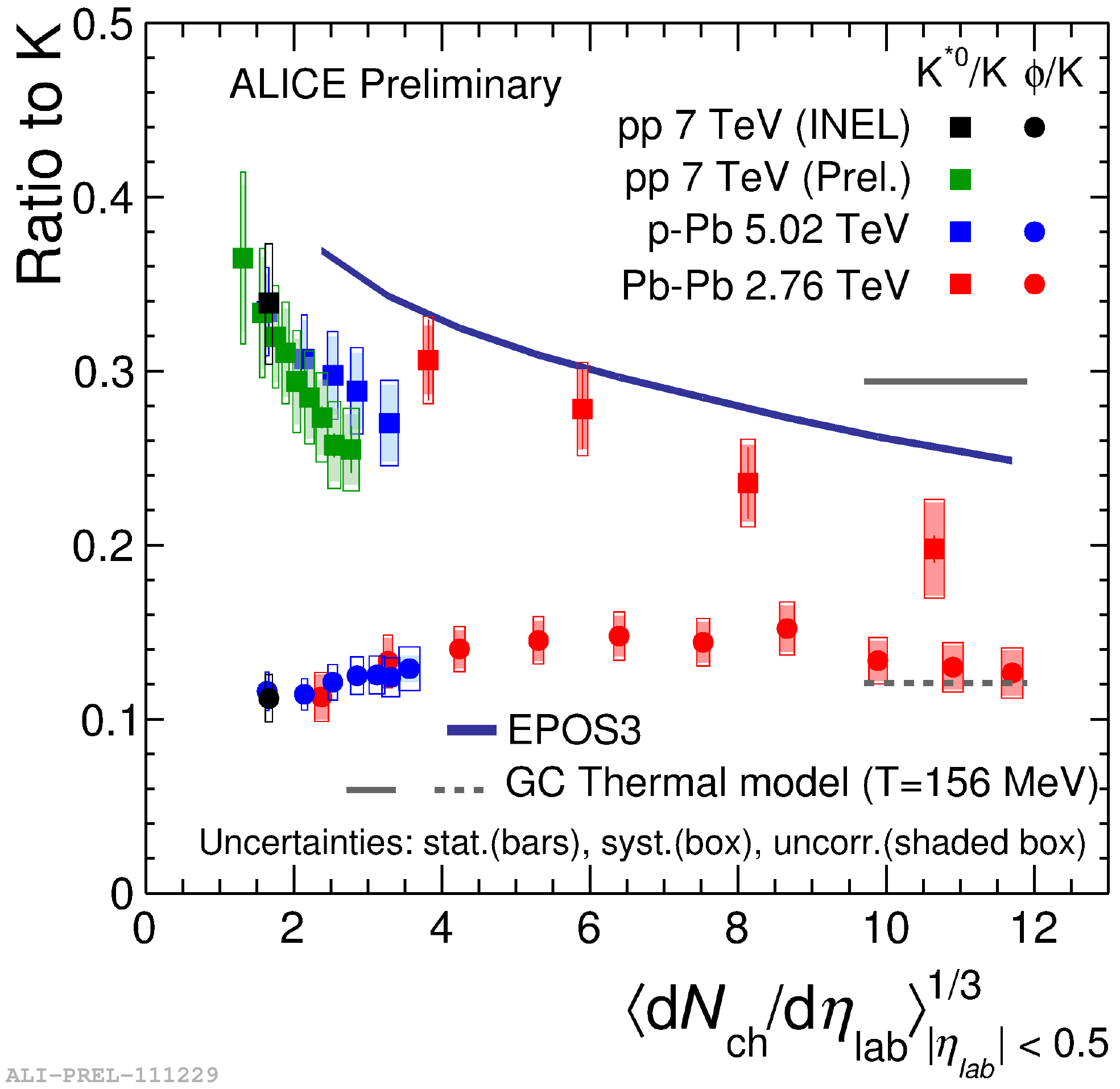}
\caption{Ratios \rhpix, \ksk, and \phik~\cite{ALICE_Kstar_phi_PbPb,ALICE_Kstar_phi_pPb5,ALICE_kstar_phi_7TeV} as functions of \dncr for various collision systems, with grand-canonical thermal-model~\cite{THERMUS,Stachel_SQM2013} and EPOS~\cite{EPOS_resonances_PbPb} calculations.}
\label{fig1}
\end{figure}

The \rhm, \ksm, and \phm mesons (hereafter \rh, \ks, and \ph) and the \sigs and \xs baryons are measured using invariant-mass analyses to reconstruct their hadronic decays ($\rh\rightarrow\pim\pip$, $\ks\rightarrow\pix\mathrm{K}^{\mp}$, $\ph\rightarrow\km\kp$, $\sigs\rightarrow\Lambda\pix$, and $\xs\rightarrow\Xi^{-}\pip$).  For the baryonic resonances, the intermediate decay daughters $\Lambda$ and $\Xi^{-}$ are identified through selections based on their masses and decay topologies.  Combinatorial backgrounds, constructed using either like-charge pairs or event mixing, are subtracted from the distributions of unlike-charge pairs for multiple \ptt and centrality/multiplicity intervals.  The resulting invariant-mass distributions are then fitted.  For the \rh, a cocktail is used, including a smooth function to describe a continuum residual background, plus peaks accounting for the contributions of the $K^{0}_{\mathrm{S}}$, $\omega(782)$, \ks, f$_{0}(980)$, and f$_{2}(1270)$.  The shape of the \rh peak is described by the product of a relativistic $p$-wave Breit-Wigner function, a phase-space factor, a mass-dependent reconstruction efficiency, and a S\"{o}ding interference term~\cite{Beddall_Soeding}.  For the other resonances, the background-subtracted invariant-mass distributions are fitted using Breit-Wigner or Voigtian peaks added to first- or second-order polynomials to describe the residual backgrounds; see~\cite{ALICE_Kstar_phi_PbPb,ALICE_Sigmastar_Xistar_pp7} for further details.  The \ptt spectra, which are corrected for efficiency, acceptance, and branching ratios, are fitted using L\'{e}vy-Tsallis functions (for pp and \ppb collisions) or Boltzmann-Gibbs blast-wave functions (for \pb collisions) so that the total integrated yields and mean transverse momenta \mpt can be extracted.  The \ptt spectra and yields of the \rh have been measured in pp and \pb collisions at \rsnn[2.76~TeV].  The \ks and \ph have been measured in pp, \ppb, and \pb collisions at various energies in different multiplicity or centrality intervals.  The \sigs and \xs have been measured in pp collisions at \rs and in \ppb collisions at \rsnn[5.02~TeV].

Figure~\ref{fig1} shows the ratios of integrated yields \rhpix, \ksk, and \phik for the various collision systems~\cite{ALICE_Kstar_phi_PbPb,ALICE_Kstar_phi_pPb5,ALICE_kstar_phi_7TeV}.  A centrality-dependent suppression of the \rhpix and \ksk ratios has been observed in \pb collisions.  These ratios in central \pb collisions are also suppressed with respect to grand-canonical thermal-model calculations~\cite{THERMUS,Stachel_SQM2013} with a chemical freeze-out temperature of 156~MeV.  This suppression may be the result of re-scattering of the \rh and \ks decay products in the hadronic phase of the medium (re-scattering being dominant over regeneration of these resonances) and is at least qualitatively reproduced by calculations using the EPOS model~\cite{EPOS_resonances_PbPb}.  In contrast, the \ph lives 10 times longer than the \ks and 35 times longer than the \rh; it decays predominantly after the end of the hadronic phase and is not significantly affected by re-scattering or regeneration.  Furthermore, an apparent multiplicity-dependent suppression of the \ksk ratio has been observed in pp and \ppb collisions, which might be an indication of a hadron-gas phase with non-zero lifetime in high-multiplicity pp and \ppb collisions.  The \ks and \ph mesons have also been measured in inelastic pp collisions at \rs[13~TeV].  No energy evolution in the \ptt-integrated \ksk and \phik ratios is observed from RHIC to the top LHC energy.  Furthermore, the \sigsl and \xsx ratios in \ppb collisions are consistent with the value previously measured by ALICE in minimum-bias pp collisions at \rs[7~TeV]~\cite{ALICE_Sigmastar_Xistar_pp7}.  No multiplicity dependence is observed for these ratios in \ppb collisions.  This implies that the $\sigs/\pion$ ratio has the same evolution with multiplicity as the $\Lambda/\pion$ ratio, and similarly that $\xs/\pion$ evolves like the $\Xi^{-}/\pion$ ratio.  The $\Lambda/\pion$ and $\Xi^{-}/\pion$ ratios are enhanced as functions of multiplicity in \ppb collisions~\cite{ALICE_multistrange_pPb}; the evolution of the $\sigs/\pion$ and $\xs/\pion$ suggests that the enhancement depends on their strangeness content and not on their masses.

\begin{figure}
\includegraphics[width=8cm]{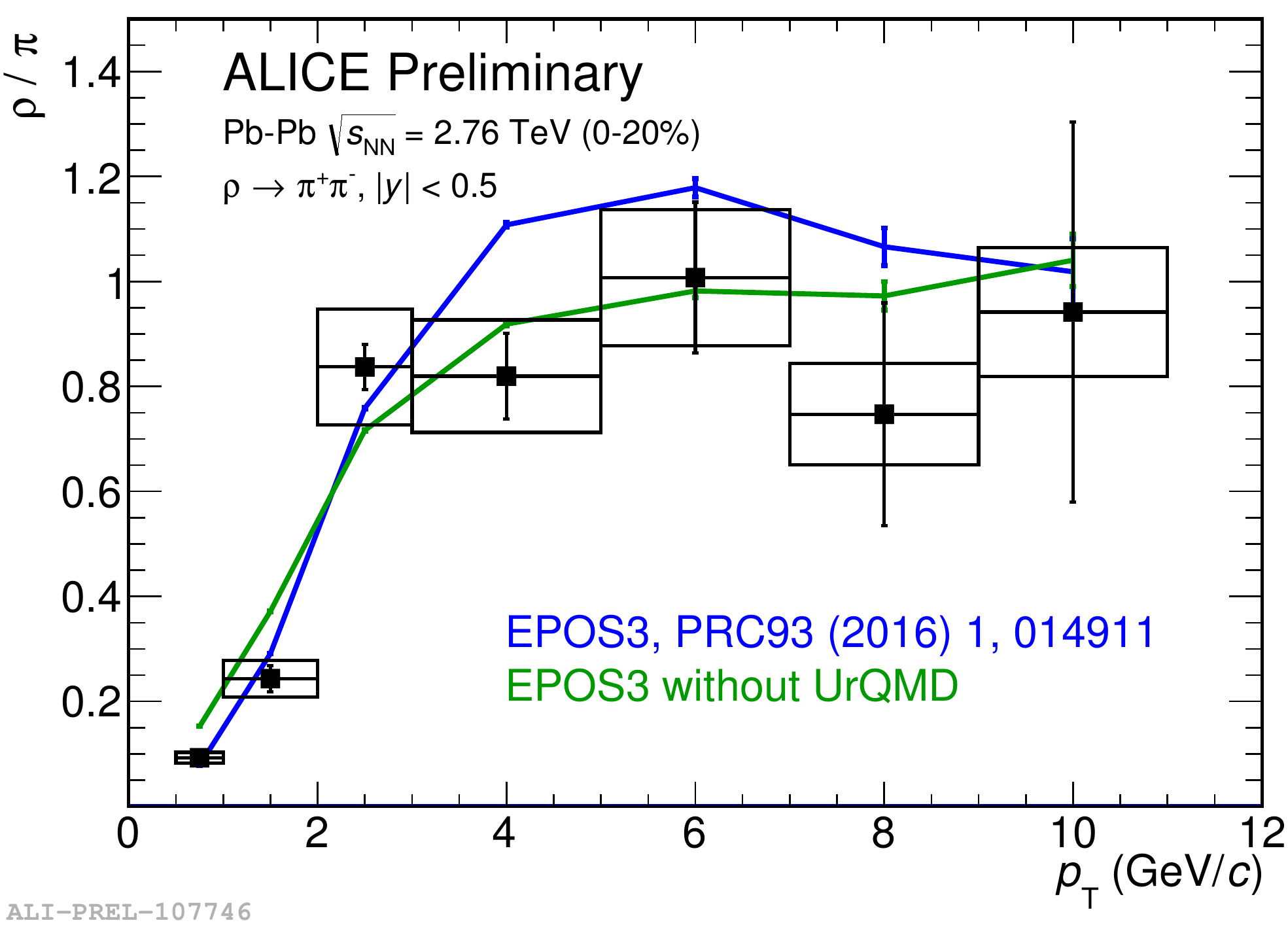}
\includegraphics[width=6.5cm]{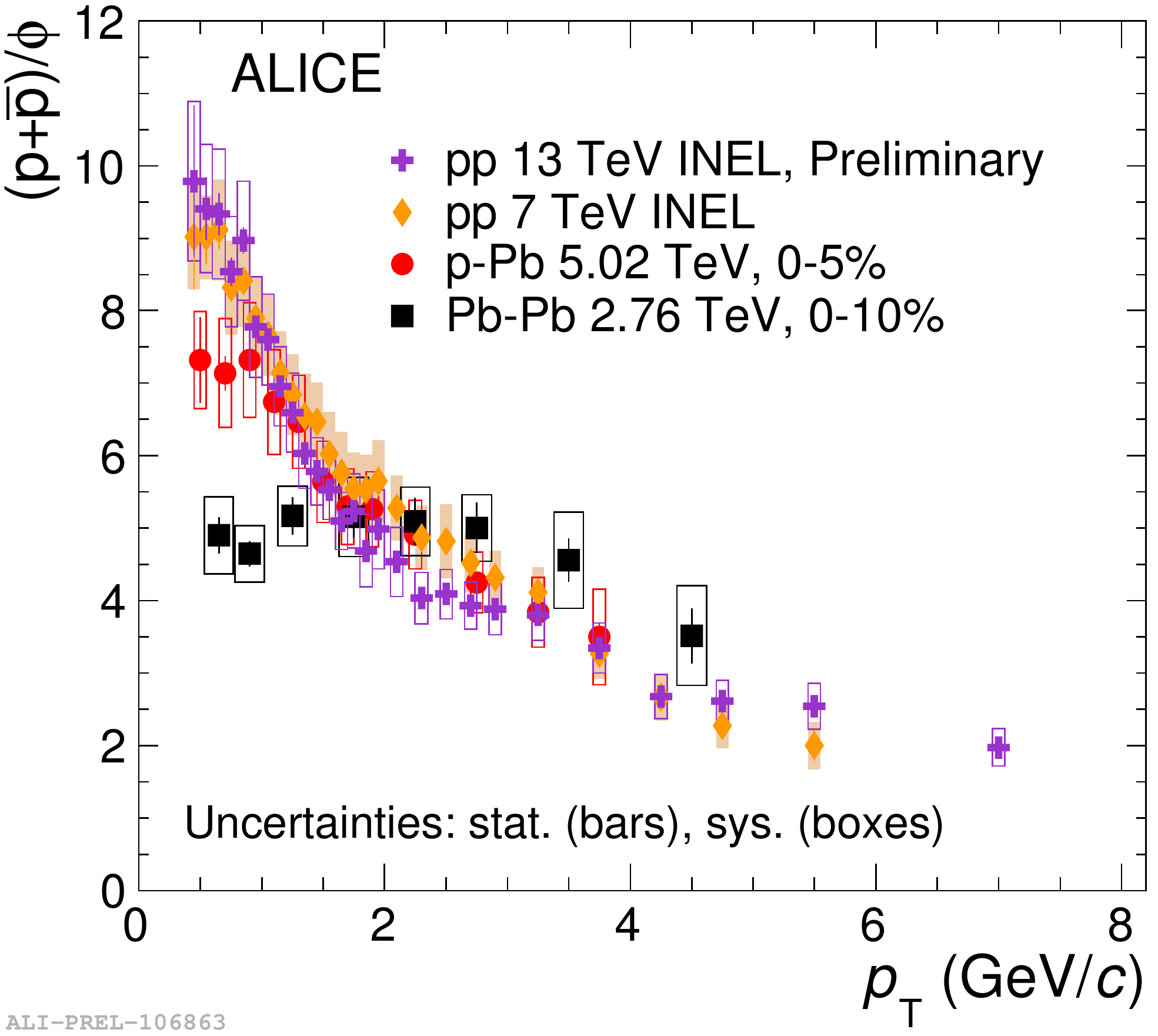}
\caption{Ratios \rhpi and \pphi as functions of \ptt for various collision systems~\cite{ALICE_Kstar_phi_PbPb,ALICE_Kstar_phi_pPb5,ALICE_kstar_phi_7TeV}.  EPOS calculations~\cite{EPOS_resonances_PbPb} of the \rhpi ratio are also shown.}
\label{fig2}
\includegraphics[width=14.875cm]{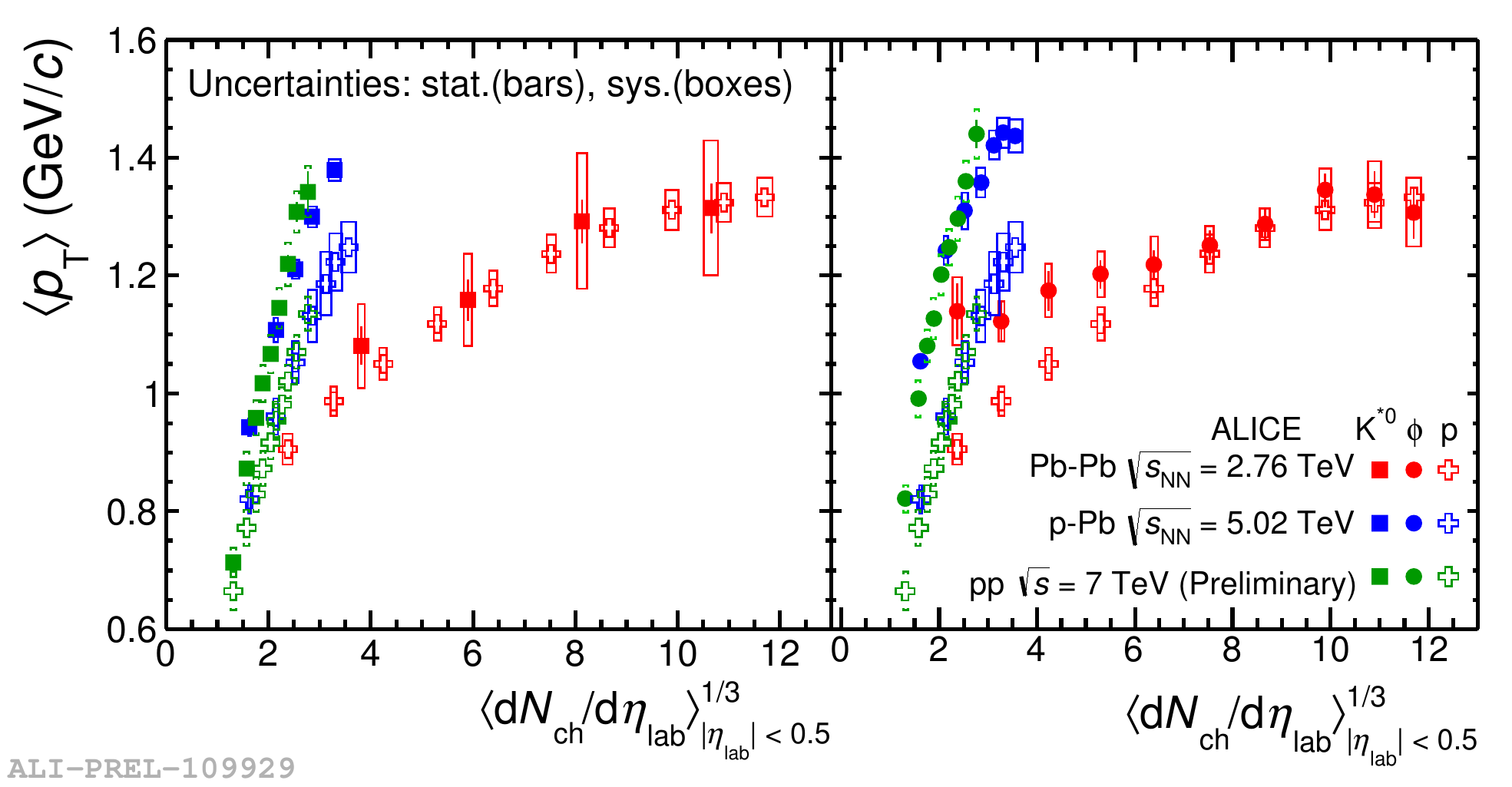}
\caption{Mean transverse momentum \mpt values of \ks, p, and \ph in various collision systems~\cite{ALICE_Kstar_phi_PbPb,ALICE_Kstar_phi_pPb5,ALICE_piKp_PbPb,ALICE_piKp_pPb}.}
\label{fig3}
\end{figure}

The \ptt dependent \rhpi ratio has also been calculated in \pb collisions, see Fig.~\ref{fig2}(a).  In central collisions, this ratio is fairly well described by EPOS~\cite{EPOS_resonances_PbPb} with its UrQMD module turned on, but is overestimated by at least 30\% for $\ptt<2$~\gvc if UrQMD is turned off.  Both EPOS calculations describe the ratio in peripheral \pb collisions well.  This suggests that the hadronic phase, which includes re-scattering effects in UrQMD, may indeed be responsible for the observed suppression of the \rh yield in central \pb collisions.

Baryon-to-meson ratios (\textit{e.g.}, p/\pion and $\Lambda$/K$^{0}_{S}$) have been observed to be enhanced in \pb collisions with respect to pp collisions at intermediate \ptt ($1.5\lesssim\ptt\lesssim 6$~\gvc)~\cite{ALICE_piKp_PbPb,ALICE_k0s_Lambda_PbPb}.  This enhancement may be attributed to hadronization by recombination~\cite{Fries_Muller_2003,Coalescence_Review_2008} or to hydrodynamic flow that modifies the particle \ptt spectra~\cite{VISH2p1_MCGlb,VISH2p1_MCKLN,KRAKOW}.  The \pphi ratio, in which the baryon numerator and meson denominator particles have very similar masses, can be used to further study these effects.  This ratio is constant in central \pb collisions for $\ptt<4$~\gvc~\cite{ALICE_Kstar_phi_PbPb}, but becomes sloped for peripheral \pb, \ppb, and pp collisions, as shown in Fig.~\ref{fig2}(b).  The constant behavior in central \pb collisions is consistent with the basic hydrodynamic assumption, although the behavior is also reproduced by some recombination models~\cite{Minissale_2015}.

The dependence of the shapes of \ptt spectra on mass and quark content can also be studied using the mean transverse momentum \mpt.  The \mpt values of the \ks, p, and \ph (which all have similar masses) are shown in Fig.~\ref{fig3} for different collision systems~\cite{ALICE_Kstar_phi_PbPb}.  For central \pb collisions, the \mpt values for these three particles are consistent.  This suggests that the shapes of the \ptt spectra are determined primarily by the particle masses, which would be expected for hydrodynamic behavior.  However, this behavior is not observed for smaller collision systems, where the proton is observed to have lower \mpt values than the two mesonic resonances.  The \mpt values in pp and \ppb collisions also follow different trends and rise faster with multiplicity than in \pb collisions.  In pp and \ppb collisions, the \mpt values approach or even exceed the values measured in central \pb collisions.

In summary, the short lifetimes of hadronic resonances make them useful probes for the study of the hadronic phase of heavy-ion collisions.  Resonances can also be used along with stable hadrons in studies of the mass and quark-content dependence of mechanisms that influence the shapes of particle \ptt spectra.  The ALICE Collaboration has measured a centrality-dependent suppression of the \rhpi and \ksk ratios in \pb collisions which can be described by EPOS calculations with UrQMD.  Multiplicity-dependent suppression of the \ksk ratio is also observed in pp and \ppb collisions.  In contrast, the \phik ratio does not exhibit a centrality- or multiplicity-dependent suppression.  This behavior may be due to the loss of the \rh and \ks signals caused by re-scattering of their decay products in the hadronic phase, while the \ph decays mostly after the hadronic phase and is not affected.  The \sigsl and \xsx ratios are not observed to depend on system size or activity in pp and \ppb collisions.  Comparisons of resonance and proton \mpt values suggest that hydrodynamics can describe the \ptt spectra of those particle in central \pb collisions, while other effects may be necessary to describe the \ptt spectra in smaller collision systems.  The ALICE Collaboration is currently extending these measurements to higher energies, other resonances ($\Lambda$(1520) and $K^{*}(892)^{\pm}$), and other collision systems (\textit{e.g.}, baryonic resonances in \pb collisions).


\section*{References}
\begin{thebibliography}{9}

\bibitem{Bleicher_Stoecker}
Bleicher M and St\"{o}cker H 2004 {\em J. Phys.\/} G {\bf 30} S111--8

\bibitem{Markert_thermal}
Markert C \akea 2002 {\em AIP Conf. Proc.\/} \textbf{631} 533--52
  
\bibitem{Vogel_Bleicher}
Vogel S and Bleicher M 2005 {\em Ricerca Scientifica ed Educazione Permanente\/} Supplemento N. 124 ed I Iori and A Bortolotti 
(Milan: Universit\`{a} degli Studi di Milano) pp 116--9 (\textit{Preprint} nucl-th/0505027)

\bibitem{Torrieri_thermal}
Torrieri G and Rafelski J 2001 {\em Phys. Lett.\/} B {\bf 509} 239--45

\bibitem{Torrieri_thermal_2001b}
Rafelski J \akea 2001 {\em Phys. Rev.\/} C {\bf 64} 054907

\bibitem{ALICE_Kstar_phi_PbPb}
Abelev B \akea~(ALICE Collaboration) 2015 {\em Phys. Rev.\/} C {\bf 91} 024609

\bibitem{ALICE_Kstar_phi_pPb5}
Adam J \akea~(ALICE Collaboration) 2016 {\em Eur. Phys. J.\/} C {\bf 76} 245

\bibitem{ALICE_kstar_phi_7TeV}
Abelev B \akea~(ALICE Collaboration) 2012 {\em Eur. Phys. J.\/} C {\bf 72} 2183

\bibitem{THERMUS}
Wheaton S, Cleymans J and Hauer M 2009 {\em Comput. Phys. Commun.\/} {\bf 180} 84-106

\bibitem{Stachel_SQM2013}
Stachel J, Andronic A, Braun-Munzinger P and Redlich K 2014 {\em J. Phys.: Conf. Ser.\/} {\bf 509} 012019

\bibitem{EPOS_resonances_PbPb}
Knospe A G~\akea 2016 {\em Phys. Rev.\/} C {\bf 93} 014911

\bibitem{Beddall_Soeding}
Beddall A, Beddall A and Bing\"{u}l A 2008 {\em Acta Phys. Polon.\/} B {\bf 39} 173-80

\bibitem{ALICE_Sigmastar_Xistar_pp7}
Abelev B \akea~(ALICE Collaboration) 2015 {\em Eur. Phys. J.\/} C {\bf 75} 1

\bibitem{ALICE_multistrange_pPb}
Adam J \akea~(ALICE Collaboration) 2016 {\em Phys. Lett.\/} B {\bf 758} 389-401

\bibitem{ALICE_piKp_PbPb}
Abelev B \akea~(ALICE Collaboration) 2013 {\em Phys. Rev.\/} C {\bf 88} 044910

\bibitem{ALICE_piKp_pPb}
Abelev B \akea~(ALICE Collaboration) 2013 {\em Phys. Lett.\/} B {\bf 728} 25-38

\bibitem{ALICE_k0s_Lambda_PbPb}
Abelev B \akea~(ALICE Collaboration) 2013 {\em Phys. Rev. Lett.\/} {\bf 111} 222301

\bibitem{Fries_Muller_2003}
Fries R \akea 2003 {\em Phys. Rev. Lett.\/} {\bf 90} 202303

\bibitem{Coalescence_Review_2008}
Fries R \akea 2008 {\em Annu. Rev. Nucl. Part. Sci\/} {\bf 58} 177--205

\bibitem{VISH2p1_MCGlb}
Qui Z \akea 2012 {\em Phys. Lett.\/} B {\bf 707} 151--5

\bibitem{VISH2p1_MCKLN}
Shen C \akea 2011 {\em Phys. Rev.\/} C {\bf 84} 044903

\bibitem{KRAKOW}
Bo\.{z}ek P and Wyskiel-Piekarska I 2012 {\em Phys. Rev.\/} C {\bf 85} 064915

\bibitem{Minissale_2015}
Minissale V \akea 2015 {\em Phys. Rev.\/} C {\bf 92} 054904

\end{thebibliography}
\end{document}